\documentclass[conference,final,twocolumn,10pt,twoside]{IEEEtran}
\usepackage{layout}
\IEEEoverridecommandlockouts 

\ifCLASSINFOpdf

\else
\fi

\usepackage{mathtools}
\usepackage{amsmath}
\usepackage{amssymb}
\usepackage{cite}
\usepackage{graphicx}
\usepackage{siunitx}
\DeclareMathOperator{\E}{\mathbb{E}}
\begin{document}
%
\title{Performance Analysis of Multi-Hop Underwater Wireless Optical Communication Systems (Extended Version)}
\author{\IEEEauthorblockN{Mohammad Vahid Jamali, Ata Chizari, and Jawad A. Salehi}
\IEEEauthorblockA{Department of Electrical Engineering, Sharif University of Technology, Tehran, Iran\\Email: m\_v\_jamali@ee.sharif.edu, chizari.ata@ee.sharif.edu, jasalehi@sharif.edu
}}



\maketitle
\begin{abstract}
In this paper, we evaluate the end-to-end bit error rate (BER) of point-to-point underwater wireless optical communication (UWOC) systems with multi-hop transmission. To do so, we analytically derive the BER expression of a single-hop UWOC link as the building block for end-to-end BER evaluation. We also apply photon-counting method to evaluate the system BER in the presence of shot noise. Moreover, we use Gauss-Hermite quadrature formula to obtain the closed-form solutions for the system BER in the case of log-normal underwater fading channels. Our analytical treatment involves all the impairing effects of the underwater optical channel, namely absorption, scattering and fading. Numerical results demonstrate that multi-hop transmission by alleviating the aforementioned impairing effects of the channel, can significantly improve the system performance and extend the viable end-to-end communication distance. 
For example, dual-hop transmission in $22.5$ m and $45$ m coastal water links can provide $17.5$ dB and $39$ dB performance enhancement at the BER of $10^{-6}$, respectively.
\end{abstract}

\IEEEpeerreviewmaketitle

\section{Introduction}
Nowadays the growing interest to underwater explorations necessitates design of appropriate and efficient underwater communication methods and systems. In comparison to the traditional underwater communication method, namely acoustic communication, the optical counterpart has three interesting advantages: higher bandwidth, lower time latency and higher security. These unique features make underwater wireless optical communication (UWOC) as a powerful alternative for high speed and large data underwater communications. However, presently UWOC systems have the capability to communicate through ranges that are typically less than $100$ m, which hampers their widespread usage. This impediment is mainly due to the three degrading effects in UWOC channels, i.e., absorption, scattering and turbulence which cause loss, inter-symbol interference (ISI) and fading on the propagating light wave, respectively.

Although, many worthwhile researches have been carried out to investigate the performance of free-space optical (FSO) communication systems over turbulent atmosphere channels \cite{safari2008relay,navidpour2007ber,andrews2001laser,zhu2002free}, study and design of appropriate UWOC systems yet have received less attention. The primary works in UWOC area have mainly focused on investigating the absorption and scattering effects of underwater optical channels \cite{petzold1972volume,mobley1994light}. Meanwhile, UWOC channel impulse response has been modeled using Monte Carlo (MC) approach in \cite{tang2014impulse}. Also a cellular UWOC network based on optical code division multiple access (OCDMA) technique has been proposed in \cite{akhoundi2015cellular} while potential applications and challenges
of such a network is elaborated in \cite{akhoundi2016cellular}. Furthermore, beneficial application of serial relaying on the performance of OCDMA-based underwater users is investigated in \cite{jamali2015ocdma}.

 On the other hand, turbulence-induced fading in UWOC channels has been studied in \cite{nikishov2000spectrum,korotkova2012light}. Moreover, the average BER of an UWOC system with respect to the log-normal distribution for fading statistics has been evaluated in \cite{gerccekciouglu2014bit,yi2015underwater}. Also in \cite{jamali2015performanceMIMO} the authors have proposed spatial diversity technique, i.e., multiple-input multiple-output transmission over UWOC links, to mitigate degrading effects of turbulence-induced fading; and therefore to improve the system performance.

The research in this paper is inspired by the need to design an UWOC system to support longer ranges communications with realistic transmit powers. It has been shown that all the above mentioned impairing effects of UWOC channels are incremental functions of distance \cite{tang2014impulse,korotkova2012light}. This distance dependency motivates us to design some intermediate nodes, namely relay nodes, in order to divide a long communication distance to shorter ones each with much reduced absorption, scattering and fading effects. Therefore, in this paper we consider multi-hop transmission over turbulent underwater  optical channel. Relay nodes operate based on bit detect-and-forward (BDF) strategy, i.e., each intermediate node only detects the received signal from the previous node and forwards the detected signal to the next node. 

The rest of the paper is organized as follows. In Section II, the channel and system models are described. In Section III, we analytically derive the BER expressions of a single-hop UWOC link as the building block for end-to-end BER evaluation. We also apply photon-counting method to evaluate the system BER in the presence of shot noise. In section IV, we use the results of Section III to analyze the end-to-end performance of multi-hop UWOC systems. In Section V, we provide the numerical results for our derived analytical expressions as well as MC simulations. Finally, we conclude the paper in Section VI.
\section{Channel and System Model}
\subsection{Channel Model}
As discussed in the previous section, UWOC channel imposes absorption, scattering and turbulence on the propagating light wave. In order to include the absorption and scattering effects, we apply Monte Carlo method to simulate the UWOC channel impulse response similar to \cite{tang2014impulse} and \cite{cox2012simulation}. We denote this fading-free impulse response of the $i$th hop by $h_0^{(i)}(t)$.

On the other hand, to take into account the $i$th hop turbulence effects we multiply $h_0^{(i)}(t)$ by a fading coefficient ${\tilde{h}}^{(i)}$, which for weak oceanic turbulence can be modeled as a random variable (RV) with log-normal probability density function (PDF) \cite{gerccekciouglu2014bit,yi2015underwater} as; 
\begin{align} \label{pdf lognormal}
\!\!\!f_{{\tilde{h}}^{(i)}}\!({\tilde{h}}^{(i)} )=\frac{1}{2{{\tilde{h}}^{(i)}}\sqrt{2\pi {\sigma }^2_{X_{i}}}}{\rm exp}\!\left(\!\!-\frac{{\left({{\rm ln}  ({{\tilde{h}}^{(i)}})\ }\!\!\!-\!2{\mu }_{X_{i}}\right)}^2}{8{\sigma }^2_{X_{i}}}\!\right)\!\!,
 \end{align}
where  ${\mu }_{X_{i}}$ and ${\sigma }^{2}_{X_{i}}$ are respectively the mean and variance of the Gaussian distributed log-amplitude factor $X_{i}=\frac{1}{2}{\rm ln}({{\tilde{h}}^{(i)}})$. To guarantee that fading neither attenuates nor amplifies the average power, we normalize fading coefficients as $\E[{\tilde{h}}^{(i)}]=1$, which implies that ${\mu }_{X_{i}}=-{\sigma }^{2}_{X_{i}}$.

For light wave with instantaneous intensity $I_{i}$, scintillation index is defined as ${\rm S.I.}=\left({\E[{I^2_{i}}]-\E^2[{I_{i}}]}\right)/{\E^2[{I_{i}}]}$, which is thoroughly studied for optical plane and spherical waves propagating in underwater turbulent channel \cite{korotkova2012light}. It can be shown that for weak optical turbulence, scintillation index relates to the log-amplitude variance as ${\rm S.I.}={\rm exp}(4{\sigma}^2_{X_{i}})-1$. 
\subsection{System Model}
\begin{figure}
     \centering
     \includegraphics[width=3.4in]{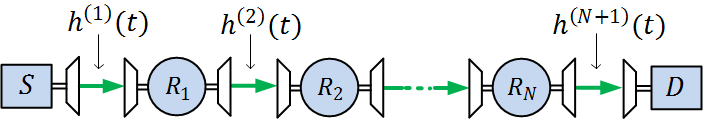}
     \caption {Block diagram of the UWOC system with multi-hop transmission.}
\end{figure}
As it is illustrated in Fig. 1, we consider an UWOC system with $N$ intermediate relay nodes in a serial topology. Each of the relay nodes operates based on BDF strategy, i.e., each $i$th node\footnote{In our system model, we denote the source node $S$ as $0$th node, relay $R_i$ as $i$th node, and destination $D$ as ($N+1$)th node.} after detecting the received optical signal, transmits the detected signal with the average transmitted power per bit of $P_b^{(i)}$. Therefore, the transmitted data sequence of the $i$th node has the form $
S^{(i)}\left(t\right)=\sum^{\infty }_{k=-\infty }{b_k^{(i)}P^{(i)}\left(t-kT_b\right)}$,
in which $T_b$ is the bit duration time and $b_k^{(i)}\in \left\{0,1\right\}$ is the $k$th time slot transmitted bit of the $i$th node, representing on-off keying (OOK) modulation. $P^{(i)}(t-kT_b)$ is the $k$th time slot transmitted optical signal of the $i$th node, which has the average power of $P_b^{(i)}$. For the sake of fairness, we assume that the total transmitted power per bit of multi-hop transmission is equal to the transmitted power per bit of single-hop transmission, i.e., $\sum_{i=0}^{N}P_b^{(i)}=P_b$. 

The received optical signal of the $i$th node after propagating through underwater optical channel with aggregated impulse response of $h^{(i)}(t)={\tilde{h}}^{(i)}h_0^{(i)}(t)$ can be expressed as;
\begin{equation} \label{y^i}
y^{(i)}(t)=S^{(i-1)}(t)*h^{(i)}(t)={\tilde{h}}^{(i)}\!\!\sum^{\infty}_{{k=-\infty}}\!\!\!{b^{(i-1)}_k\Gamma^{(i)} \left(t-kT_b\right)},
\end{equation}
where $\Gamma^{(i)}(t)=P^{(i-1)}(t)*h_0^{(i)}(t)$, and $*$ denotes the convolution operation. Each receiver, either within the relay nodes or in the destination performs symbol-by-symbol processing, which is suboptimal in the presence of ISI \cite{einarsson1996principles}. In other words, each receiver integrates over each $T_b$ seconds and compares the result with an appropriate threshold to detect the received optical signal. It is worth noting that various noise components, namely signal-dependent shot noise, background light, dark current and thermal noise, all affect the aforementioned detection process.
\section{BER of Single-Hop UWOC System}
In this section, we analytically obtain the average BER of single-hop transmission. We also apply photon-counting method \cite{einarsson1996principles} to evaluate the $i$th node conditional BER in the presence of shot noise.
\subsection{Analytical Approach}
As discussed above, various noise components, i.e., signal-dependent shot noise, background light, dark current and thermal noise all affect the system performance. Since these components are additive and independent from each other, in this subsection we model them as an equivalent noise component with Gaussian distribution \cite{lee2004part}. Moreover, we assume that the signal-dependent shot noise is negligible with respect to the other noise components and hence the noise variance is independent from the incoming optical signal power \cite{jamali2015ber}. Based on Eq. \eqref{y^i}, the $0$th time slot integrated current of the receiver output can be expressed as;
\begin{align} \label{eq4}
r^{(b_0)}=b_0{\tilde{h}}\gamma^{(s)}+{\tilde{h}}\sum_{k=-L}^{-1}b_k\gamma^{(I,k)}+v_{T_b},
\end{align}
where $\gamma^{(s)}=\boldmath{R}\int_{0}^{T_b}\Gamma(t)dt$, $\gamma^{(I,k)}=\boldmath{R}\int_{-kT_b}^{-(k-1)T_b}\Gamma(t)dt$, $\boldmath{R}={\eta q}/{hf}$ is the photodetector responsivity, $\eta$ is the photodetector quantum efficiency, $q=1.602\times10^{-19}$ C is electron charge, $h=6.626\times10^{-34}~{\rm J/s}$ is Planck's constant, $f$ is the optical source frequency and $L$ denotes the channel memory. It is worth noting that $\gamma^{(I,k\neq0)}$ refers to the ISI effect and $\gamma^{(I,k=0)}$ interprets the desired signal contribution, i.e., $\gamma^{(I,k=0)}=\gamma^{(s)}$. Furthermore, $v_{T_b}$ is the receiver integrated noise component, which has a Gaussian distribution with mean zero and variance $\sigma^2_{T_b}$ \cite{lee2004part,jaruwatanadilok2008underwater}. 

Assuming the availability of channel state information (CSI), the receiver compares its integrated current over each $T_b$ seconds with threshold value of ${\tilde{h}}\gamma^{(s)}/2$. Therefore, the conditional probability of error when ``$b_0$" is sent, can be obtained as;
\begin{align} \label{eq5}
P_{be|b_0,\tilde{h},b_k}\!\!\!=\!Q\!\!\left(\!\frac{{\tilde{h}}\!\left[\gamma^{(s)}\!\!+\!(-1)^{b_0+1}\!\sum_{k=-L}^{-1}2b_k\gamma^{(I,k)}\!\right]}{2\sigma_{T_b}}\!\right),
\end{align}
where $Q\left(x\right)=({1}/{\sqrt{2\pi }})\int^{\infty }_x{{\rm exp}({-{y^2}/{2}})}dy$ is the Gaussian-Q function. The final BER can be obtained by averaging the conditional BER over fading coefficient $\tilde{h}$ and all $2^L$ possible data sequences for $b_k$s, as follows;
\begin{align}\label{eq7}
P_{be}=\frac{1}{2^L}\sum_{b_k}\int_{0}^{\infty}\frac{1}{2}\left[P_{be|0,\tilde{h},b_k}+P_{be|1,{\tilde{h}},b_k}\right]f_{{\tilde{h}}}(\tilde{h})d{\tilde{h}}.
\end{align}


In the case of log-normal underwater fading channel, the averaging over fading coefficient in \eqref{eq7} can be effectively calculated using Gauss-Hermite quadrature formula (GHQF) [21, Eq. (25.4.46)] as follows;
\begin{align} \label{eq9}
P_{be|b_0,b_k}\!&=\!\frac{1}{\sqrt{2\pi\sigma^2_X}}\int_{x=-\infty}^{\infty}\!\!\!\!\!\!\!\!P_{be|b_0,\tilde{h}=e^{2x},b_k}e^{-{(x-\mu_X)^2}/{2\sigma^2_X}}dx\nonumber\\
&\approx\frac{1}{\sqrt{\pi}}\sum_{q=1}^{V}w_qQ\left(C \exp\left(2x_q\sqrt{2\sigma^2_X}+2\mu_X\right)\right),
\end{align}
in which $V$ is the order of approximation, $w_q,~q=1,2,...,V$, are the weights of the $V$th order approximation and $x_q$ is the $q$th zero of the $V$th-order Hermite polynomial, $H_V(x)$ \cite{navidpour2007ber,abramowitz1970handbook}. Moreover, the parameter $C$ in \eqref{eq9} is defined as;
\begin{align}
C=\frac{\gamma^{(s)}\!\!+\!(-1)^{b_0+1}\!\sum_{k=-L}^{-1}2b_k\gamma^{(I,k)}}{2\sigma_{T_b}}.
\end{align}
\subsection{Photon-Counting Method}
In the previous subsection, we dealt with additive white Gaussian noise (AWGN) model, which considers the incoming optical signal as a constant coefficient (conditioned on $\tilde{h}$) and all the noise components as an equivalent additive Gaussian RV. This model, however is straightforward and common \cite{safari2008relay,navidpour2007ber,lee2004part}, does not consider the shot noise effect induced by the fluctuating nature of optical signals. In other words, due to the statistical nature of light as a stream of photons, in optical systems the signals have random fluctuations which can be modeled by Poisson distribution \cite{einarsson1996principles}. Also signal-dependent shot noise, dark current and background light can be modeled as Poisson distributed RVs, while thermal noise induced by the receiver electronic circuit is usual to be considered as a Gaussian distributed RV \cite{einarsson1996principles}. In this circumstances, either saddle-point or Gaussian approximations, which are based on photon-counting methods, can be used to evaluate the system BER in the presence of shot noise. In this subsection, we apply photon-counting methods to evaluate the conditional BER of the $i$th node.

 When the ($i-1$)th node transmits data bit $b_0^{(i-1)}$ on its $0$th time slot, the photo-detected count signal generated by the $i$th node integrate and dump circuit can be modeled as $u^{(i)}_{b_0^{(i-1)}}=y^{(i)}_{b_0^{(i-1)}}+v^{(i)}_{th}$,
in which $v^{(i)}_{th}$ is a Gaussian distributed RV with mean zero and variance $\sigma^2_{th}={2K_bT_rT_b}/{Rq^2}$, corresponding to the $i$th node thermal noise. $K_b$, $T_r$ and $R$ are Boltzmann's constant, the receiver
  equivalent temperature and load resistance, respectively. Moreover based on \eqref{y^i}, conditioned on $\{ b_k^{(i-1)} \}_{k=-L_i}^{-1}$ and ${\tilde{h}}^{{(i)}}$, $y^{(i)}_{b_0^{(i-1)}}$ is a Poisson distributed RV with mean;
\begin{align} \label{m^i}
m^{(i)}_{b_0^{(i-1)}}\!=\!\frac{\eta}{hf}{\tilde{h}}^{(i)}\!\!\!\!\sum^{0}_{{k=-L_i}}\!\!\!{b^{(i-1)}_k\!\!\int_{0}^{T_b}\!\!\!\Gamma^{(i)} \left(t\!-\!kT_b\right)}dt\!+\!n_{bd}^{(i)},
\end{align}
where $L_i$ is the $i$th hop channel memory and $n_{bd}^{(i)}=(n_{b}^{(i)}+n_{d}^{(i)})T_b$, in which $n_{b}^{(i)}$ and $n_{d}^{(i)}$ are the mean photoelectron count rates of background light and dark current of the $i$th node, respectively.

Now, based on saddle-point approximation, the conditional BER of the $i$th node can be calculated as \cite{einarsson1996principles};
\begin{align} \label{38}
P^{(i)}_{be|{\tilde{h}}^{{(i)}},b_k^{(i-1)}}&=\frac{1}{2}\Pr(u^{(i)}_{1}\leq\beta^{(i)})+\frac{1}{2}\Pr(u^{(i)}_{0}>\beta^{(i)}) \nonumber\\
   & =\frac{1}{2}\left[q_-(\beta^{(i)},s_1)+q_+(\beta^{(i)},s_0)\right],
 \end{align}
 where $\beta^{(i)}$ is the $i$th node threshold. $q_-(\beta^{(i)},s_1)$ and $q_+(\beta^{(i)},s_0)$ are conditional probabilities of error for ``ON" and ``OFF" states, respectively and are defined as;
   \begin{align}
   & q_-(\beta^{(i)}\!,s_1)\!=\!\frac{{\rm exp}\left[m^{(i)}_1\!\left(e^{s_1}\!-\!1\right)\!+\!{s^2_1\sigma_{th} ^2}/2\!-\!s_1\beta^{(i)}\!-\!{\rm ln}|s_1|\right]}{\sqrt{2\pi \left(m^{(i)}_1e^{s_1}+\sigma_{th} ^2+1/s^2_1\right)}},\label{39}\\
   & q_+(\beta^{(i)}\!,s_0)\!=\!\frac{{\rm exp}\left[m^{(i)}_0\!\left(e^{s_0}\!-\!1\right)\!+\!{s^2_0\sigma_{th} ^2}/2\!-\!s_0\beta^{(i)}\!-\!{\rm ln}|s_0|\right]}{\sqrt{2\pi \left(m^{(i)}_0e^{s_0}+\sigma_{th}^2+1/s^2_0\right)}}.\label{40}
   \end{align}
   It can be shown that the parameters $s_0$, $s_1$ and $\beta^{(i)}$ can be obtained by numerical solving of the following set of equations \cite{einarsson1996principles};
   \begin{align} \label{41}
    s_0: m^{(i)}_0e^{s_0}+\sigma_{th}^2s_0-\beta^{(i)}-1/{s_0}=0,
    \end{align}
    \vspace{-0.2in}
    \begin{align} \label{42}
    s_1: m^{(i)}_1e^{s_1}+\sigma_{th}^2s_1-\beta^{(i)}-1/{s_1}=0,
    \end{align}
        \vspace{-0.2in}
    \begin{align} \label{43}
    \beta^{(i)}\!:\! \frac{dP^{(i)}_{be|{\tilde{h}}^{(i)},b^{(i-1)}_k}}{d\beta ^{(i)}}=0\Rightarrow s_0q_+(\beta^{(i)},s_0)\!+\!s_1q_-\left(\beta ,s_1\right)\!=\!0.
    \end{align}

As solving the above equations may be cumbersome and time consuming, Gaussian approximation, which is simple and fast but is not as accurate as saddle-point approximation, can be used to evaluate the conditional BER based on photon-counting methods. In fact, Gaussian approximation approximates all the Poisson RVs with Gaussian distributed RVs, where each of the approximated RVs has the same mean and variance, and then leads to the following equation for the system conditional BER \cite{einarsson1996principles};
\begin{align} \label{P^i}
\!\!P^{(i)}_{\!\!be|0,{\tilde{h}}^{{(i)}}\!,b_k^{(i\!-\!1)}}\!\!=\!\!P^{(i)}_{\!\!be|1,{\tilde{h}}^{{(i)}}\!,b_k^{(i\!-\!1)}}\!\!=\!Q\!\left(\!\frac{m_1^{(i)}-m_0^{(i)}}{\!\sqrt{\!m_1^{(i)}\!\!+\!{\sigma }^2_{th}}\!+\!\sqrt{\!m_0^{(i)}\!\!+\!{\sigma }^2_{th}}}\!\right)\!\!.
\end{align}
 Hereafter, for the sake of brevity we denote the $i$th node conditional BER with $P^{(i)}_{cbe-b_0}$, which can be obtained using one of Eqs. \eqref{eq5}, \eqref{38} or \eqref{P^i}.
\section{End-to-End BER Analysis}
In this subsection, we analyze the end-to-end BER of the relay-assisted UWOC system relying on the $i$th node conditional BER, obtained in the previous section.
Let $U$ denote the number of nodes that incorrectly detect their previous node's $0$th time slot transmitted bit. For this case, the conditional end-to-end correct detection probability $P_{e2e-b_0}({\rm c}|\{\bar{\beta}\},\bar{H})$ can be obtained as;
\begin{align} \label{P_corr|}
& P_{e2e-b_0}({\rm c}|\{\bar{\beta}\},\bar{H})=\nonumber\\
&~~~\sum_{u=0}^{N+1}P_{e2e-b_0}({\rm c}|U,\{\bar{\beta}\},\bar{H})\Pr(U=u|b_0,\{\bar{\beta}\},\bar{H}),
\end{align}
in which $\bar{H}=({\tilde{h}}^{(1)},{\tilde{h}}^{(2)},...,{\tilde{h}}^{(N+1)})$ is the fading coefficients vector and $\{\bar{\beta}\}$ implies the transmitted data sequences of all transmitters. $\Pr(U=u|b_0,\{\bar{\beta}\},\bar{H})$ is the conditional probability that $u$ nodes out of $N+1$ nodes incorrectly detect the received bit. Obviously, $P_{e2e-b_0}({\rm c}|U={\rm odd},\{\bar{\beta}\},\bar{H})=0$ and $P_{e2e-b_0}({\rm c}|U={\rm even},\{\bar{\beta}\},\bar{H})=1$. Therefore, the end-to-end conditional BER can be evaluated as $P_{e2e-b_0}({\rm error}|\{\bar{\beta}\},\bar{H})=1-\sum_{u\in\Lambda_N}\Pr(U=u|b_0,\{\bar{\beta}\},\bar{H})$, where $\Lambda_N$ specifies the set of all the even numbers in the set $\{0,1,...,N+1\}$, i.e., $\Lambda_N=\{0,2,...,\left \lfloor \frac{N+1}{2} \right \rfloor\times2\}$, where $\left \lfloor x \right \rfloor$ is the integer portion of the real value $x$. 
On the other hand, $\Pr(U=u|b_0,\{\bar{\beta}\},\bar{H})$ for $u=0,1,...,N+1$ can be obtained as;
\begin{align} \label{11}
 &\Pr(U=u|b_0,\{\bar{\beta}\},\bar{H})\!=\!\sum_{s_1=1}^{N+1}\sum_{~s_2=s_1+1}^{N+1}\!\!...\!\!\sum_{s_u=s_{u-1}+1}^{N+1}\!\!\bigg(\!\!P^{(s_1)}_{cbe-b_0}\nonumber\\
& \times P^{(s_2)}_{cbe-b_0}\times...\times P^{(s_u)}_{cbe-b_0}\times \!\!\!\!\!\!\!\!\!\!\!\prod_{_{~s_{u+1}\neq s_1,s_2,...,s_u}^{~~~~~s_{u+1}=1}}^{N+1}\!\!\!\!\!\!\!\!\left[1-P^{(s_{u+1})}_{cbe-b_0}\right]\bigg).
\end{align}
Averaging $P_{e2e-b_0}({\rm error}|\{\bar{\beta}\},\bar{H})$ over $\{\bar{\beta}\}$ and $\bar{H}$ results into $P_{e2e-b_0}({\rm error})=1-\sum_{u\in\Lambda_N}\Pr(U=u|b_0)$, where $\Pr(U=u|b_0)$ is defined as;
\begin{align} \label{P(u)}
\Pr(U\!\!=\!u|b_0)\!=\!\sum_{\{\bar{\beta}\}}P(\{\bar{\beta}\})\!\int_{\bar{H}}\!\!\Pr(U\!\!=\!u|b_0,\{\bar{\beta}\},\bar{H})f(\bar{H})d\bar{H},
\end{align}
 in which $P(\{\bar{\beta}\})$ and $f(\bar{H})$ are the joint PDFs of $\bar{\beta}$s and $\tilde{h}^{(i)}$s, respectively. Since $b^{(i)}_k$s are independent with identical probability and $\tilde{h}^{(i)}$s are also independent, Eq. \eqref{P(u)} reduces to a similar form of Eq. \eqref{11} except that $P^{(i)}_{cbe-b_0}$s are replaced by $P^{(i)}_{be-b_0}$s which are the averaged form of $P^{(i)}_{cbe-b_0}$s and are defined as;
 \begin{align} \label{P_be-i}
P^{(i)}_{be-b_0}=\frac{1}{2^{L_i}}\sum_{b^{(i-1)}_k}\int_{{\tilde{h}}^{(i)}}P^{(i)}_{cbe-b_0}~f_{{\tilde{h}}^{(i)}}({\tilde{h}}^{(i)})d{\tilde{h}}^{(i)},
 \end{align}
Replacing \eqref{P_be-i} instead of $P^{(i)}_{cbe-b_0}$ in Eq. \eqref{11} results in $\Pr(U=u|b_0)$, and eventually the end-to-end average BER can be evaluated as $P_{e2e-b_0}({\rm error})=1-\sum_{u\in\Lambda_N}\Pr(U=u|b_0)$. Therefore, the end-to-end average BER of UWOC systems with serial BDF relaying can be obtained through one-dimensional integrals which in the case of weak oceanic turbulence leads to a closed form solution, using GHQF as demonstrated in \eqref{eq9}.
 
It is worth noting that in some common scenarios (e.g., the following two cases) $\Pr(U=u|b_0)$ and therefore the average BER can be expressed more explicitly.
\begin{itemize}
  \item[$\bullet$] The form of Eq. \eqref{11} suggests that $\Pr(U=u|b_0)$ decreases rapidly for larger values of $u$, since it is proportional to the product of the BERs of $u$ nodes. Therefore, to simplify the average end-to-end BER expression, we make an assumption: however, it is possible to detect a bit correctly at the destination despite of incorrect detection in some of the intermediate relay nodes, i.e., when $u\neq 0$, we neglect these fortunate events and assume that a bit can be correctly detected at the destination if and only if all the receivers detect without an error (i.e., when $u=0$). This assumption is valid particularly when each hop has a good performance, i.e., a small $P^{(i)}_{be-b_0}$. In this case, the average end-to-end BER can be evaluated as;
  \begin{align} \label{11_asli}
  P_{e2e-b_0}({\rm error})=1-\prod_{i=1}^{N+1}\left[1-P^{(i)}_{be-b_0}\right].
  \end{align}
  \item[$\bullet$] Suppose that all links have the same average BER of $P^{(i)}_{be-b_0}=P_{be-b_0}$. This is a valid assumption for example when all hops have the same link length and water quality and all receivers have the same structure. In this case, the average BER of the system can be obtained as follows;
\begin{align} \label{12}
& P_{e2e-b_0}({\rm error})=1-\nonumber\\
& \sum_{u\in\Lambda_N}\binom{N+1}{u}{\left(P_{be-b_0}\right)}^{u}\left(1-P_{be-b_0}\right)^{N+1-u}.
    \end{align}
\end{itemize}
\section{Numerical Results}
In this section, we present the numerical results for the relay-assisted UWOC system BER. We also obtain the system BER using numerical simulations to verify the accuracy of our derived analytical expressions. We simulate the channel impulse response similar to \cite{tang2014impulse,cox2012simulation} and the channel scintillation index for a propagating plane wave like \cite{korotkova2012light}. Table I shows some of the important parameters for the channel simulation and noise characterization. In our simulations we assume that the UWOC system is established in coastal water \cite{tang2014impulse}.
\begin{table}
  \centering
\caption{Some of the important parameters for the channel MC simulation and noise characterization.}
\begin{tabular}{||p{1.6in}|>{\centering\arraybackslash}p{0.3in}|>{\centering\arraybackslash}p{0.5in}||} \hline
   Coefficient & Symbol & Value\\ [0.2ex] 
   \hline \hline
  Half angle field of view & FOV &  ${40}^0$ \\ \hline 
     Receiver aperture diameter & $D_0$ & $20$ cm \\ \hline 
     Source wavelength & $\lambda $ & $532$ nm \\ \hline 
     Source full beam divergence angle & $\theta_{div}$ & $0.02^0$ \\ \hline 
     Quantum efficiency & $\eta $ & $0.8$ \\ \hline 
         Equivalent temperature & $T_e$ & $290$ K \\ \hline 
         Load resistance & $R_L$ & $100$ $\Omega$ \\ \hline 
         Dark current & $I_{dc}$ & $1.226\times {10}^{-9}$ A \\ \hline
         Background mean count rate &$n_b$& $1.8094\times10^8$ ${1/s}$ \\ \hline
         Rate of dissipation of mean-square temperature & $\chi_T$  & $2\times{{10}^{-7}}$ ${{{K^2}/{s}}}$ \\ \hline
         Rate of dissipation of turbulent kinetic energy per unit mass of fluid & $\varepsilon $ & $1.5\times{{10}^{-5}}$ ${{m^2}/{s^3}}$ \\ \hline
         Relative strength of temperature and salinity fluctuations & $w$ & $-2.5$ \\ \hline
  \end{tabular}
  \end{table}
  
 \begin{figure}
       \centering
       \includegraphics[width=3.4in]{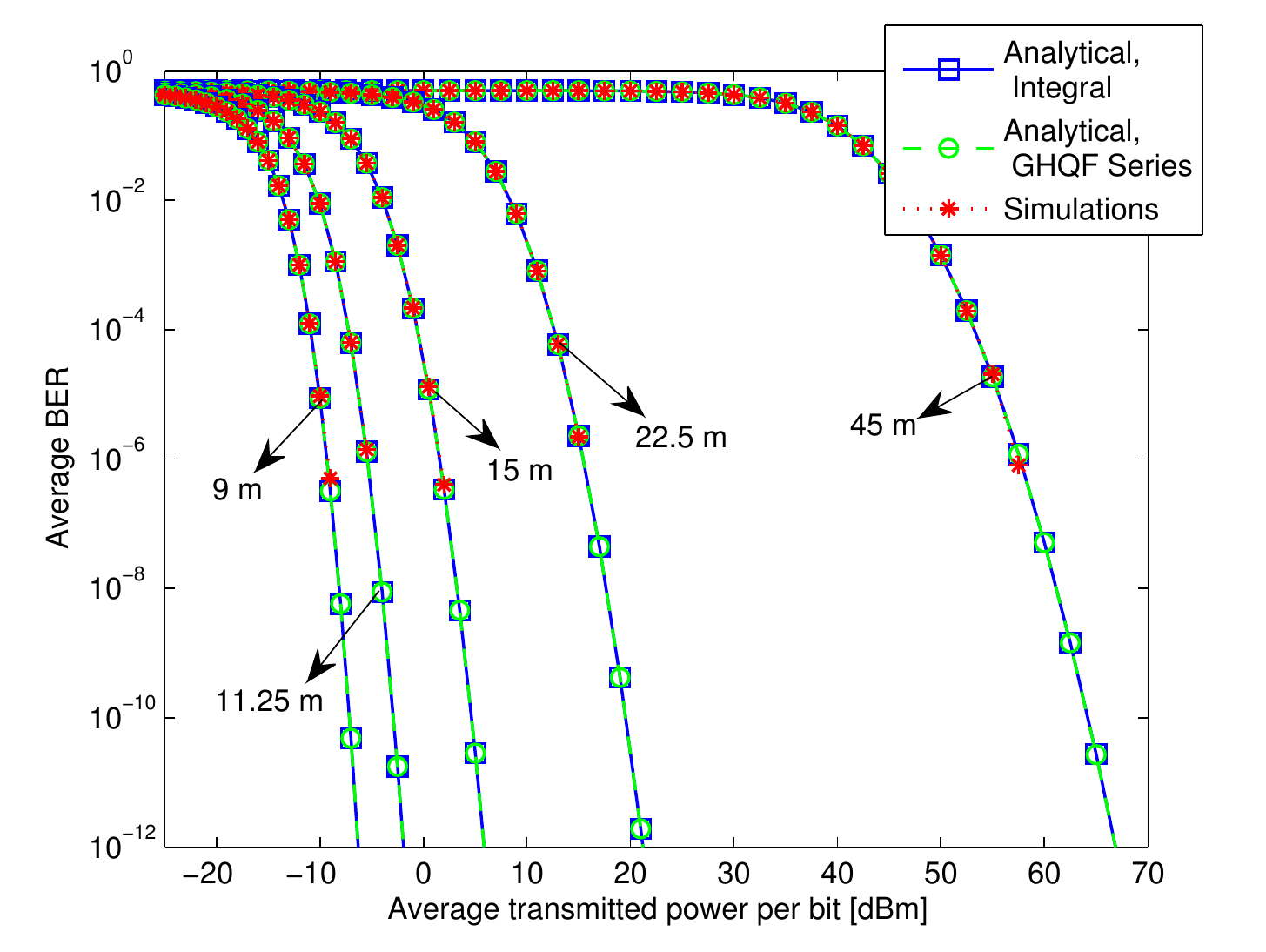}
       \caption {BER of a single-hop UWOC system with $1$ Gbps data rate and various link ranges $d_0=45$ m, $22.5$ m, $15$ m, $11.25$ m and $9$ m.}
                   \vspace{0in}
  \end{figure}
 Fig. 2 illustrates the BER of a single-hop UWOC system for various link ranges and $1$ Gbps data transmission rate, obtained using Eq. \eqref{P_be-i}. As it can be seen, increasing the range of communication severely degrades the performance. This is reasonable, since all impairing effects of the underwater channel increase rapidly with the link range. This issue motivated us in the first place, to locate some intermediate relay nodes in order to have hops with smaller ranges and therefore, with much reduced absorption, scattering and fading effects. Moreover, we applied \eqref{eq9} to calculate the system BER using GHQF. It is observed that the system BER can be effectively calculated using only $V=30$ points. Furthermore, well matches between the analytical results and numerical simulations confirm the accuracy of our derived expressions.
 
 \begin{figure}
       \centering
       \includegraphics[width=3.4in]{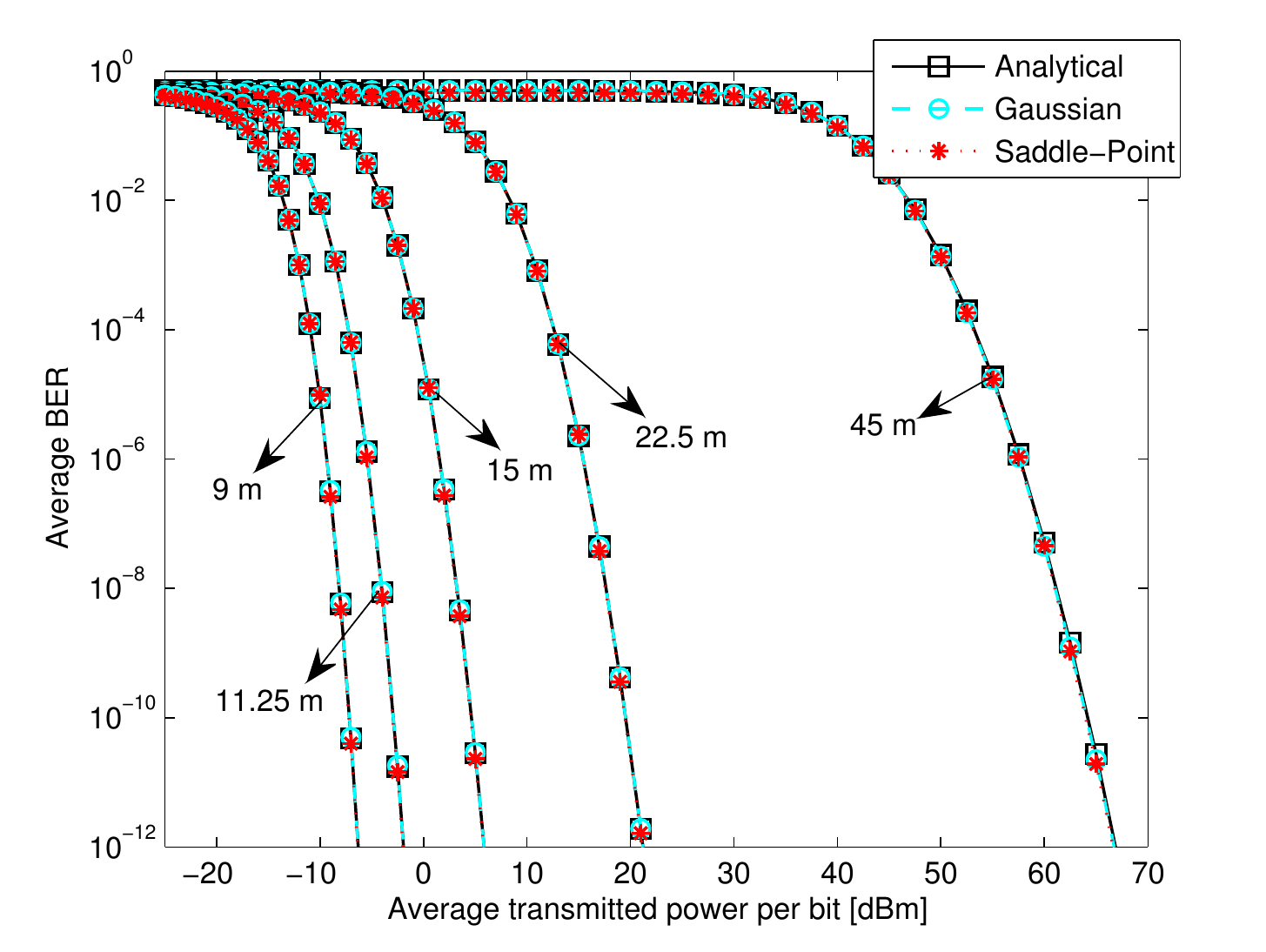}
       \caption {Comparison between different methods in evaluating the average BER of a single-hop UWOC system with $1$ Gbps data rate and different link ranges.}
  \end{figure}
 In Fig. 3, photon-counting methods, namely saddle-point and Gaussian approximations are compared. As it can be seen, Gaussian approximation can provide relatively the same results as saddle-point approximation. Hence, due to its simplicity and acceptable accuracy, Gaussian approximation can be considered as a reliable and favorable photon-counting method.
 Moreover, excellent matches between the results of analytical and photon-counting approaches further confirm the accuracy of our assumptions in Section III-A.
 
 In Fig. 4, the average BER of a relay-assisted UWOC system with $1$ Gbps data rate, $45$ m end-to-end communication distance and $N=0$, $1$, $2$, $3$ and $4$ relay nodes is obtained using Eqs. \eqref{11_asli} and \eqref{12}. As it is obvious, using intermediate relay nodes provides significant performance improvement, e.g., a dual-hop UWOC system with $45$ m and $22.5$ m end-to-end distances can introduce $39$ dB and $17.5$ dB performance enhancement at the BER of $10^{-6}$, respectively. Moreover, as it can be seen, \eqref{11_asli} provides an acceptable approximate of the system BER especially for low BERs. Only for high BER regimes a negligible discrepancy exists between the results of Eqs. \eqref{11_asli} and \eqref{12}, since \eqref{11_asli} neglects the probability of correct detection for $u\neq0$ and therefore, it provides an upper bound on the system BER.
 Comparing Figs. 2 and 3 indicates that the dominant bottleneck on the performance of a relay-assisted UWOC system is the degrading effects of each intermediate hop. In other words, a relay-assisted UWOC system with $N$ intermediate relay nodes and $d_0$ m end-to-end communication distance can achieve the performance of a single-hop UWOC system with $d_0/{(N+1)}$ m end-to-end distance and by approximately $10~{\rm log}(N+1)$ dB excess average transmitted power per bit.
 Furthermore, excellent matches between analytical and numerical simulation results confirm the correctness of our analysis for end-to-end BER characterization.
 \begin{figure}
       \centering
       \includegraphics[width=3.4in]{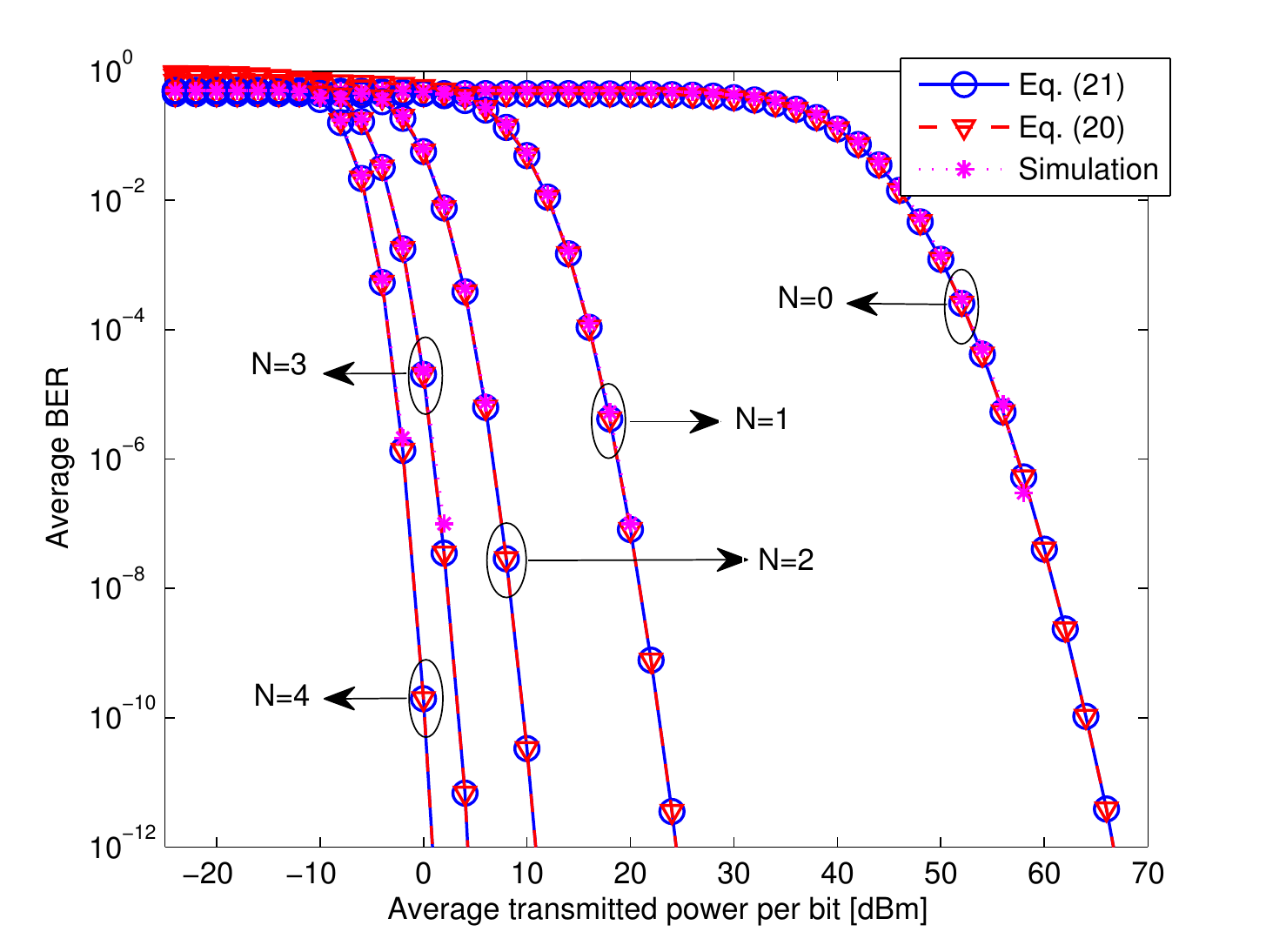}
       \caption {BER of a relay-assisted UWOC system with $1$ Gbps data rate and link range of $d_0=45$ m. $N=0$, $1$, $2$, $3$ and $4$.}
  \end{figure}

 Fig. 5 shows the BER of an UWOC system with $3$ intermediate relay nodes and with $105$ m end-to-end communication distance. First, second, third and fourth hops lengths are $22.5$ m, $25$ m, $27.5$ m and $30$ m, respectively. Since each $i$th hop has its distinct average BER $P^{(i)}_{be-b_0}$, the end-to-end BER can be obtained as $P_{e2e-b_0}({\rm error})=1-\sum_{u=0,2,4}\Pr(U=u|b_0,N=3)$,
  where $\Pr(U=u|b_0,N=3)$ can be easily achieved from Eqs. \eqref{11}-\eqref{P_be-i}.
   Also in this figure the end-to-end BER is evaluated based on Eq. \eqref{11_asli} as $P_{e2e-b_0}({\rm error})=1-\prod_{s_1=1}^{4}(1-P^{(s_1)}_{be-b_0})$, and only a few discrepancy is observed at the high BER regimes. Moreover, the effect of ISI is investigated in this figure by evaluating the end-to-end BER of the system for various data rates, i.e., $20$ Mbps, $100$ Mbps, $1$ Gbps, $5$ Gbps and $10$ Gbps, and it is observed that due to the scattering nature of UWOC channels, increasing the data rate degrades the system performance. Furthermore, this figure indicates the beneficial application of the relay-assisted topology in UWOC systems, where via $3$ intermediate relay nodes we can appropriately communicate through $105$ m coastal water with a realistic average transmitted power per bit. This available communication range significantly differs from the reported distances in the existing literature \cite{tang2014impulse,akhoundi2015cellular}. Therefore, designing the relay-assisted configuration is of utmost importance for UWOC systems, especially for longer range communications.
 \begin{figure}
       \centering
       \includegraphics[width=3.4in]{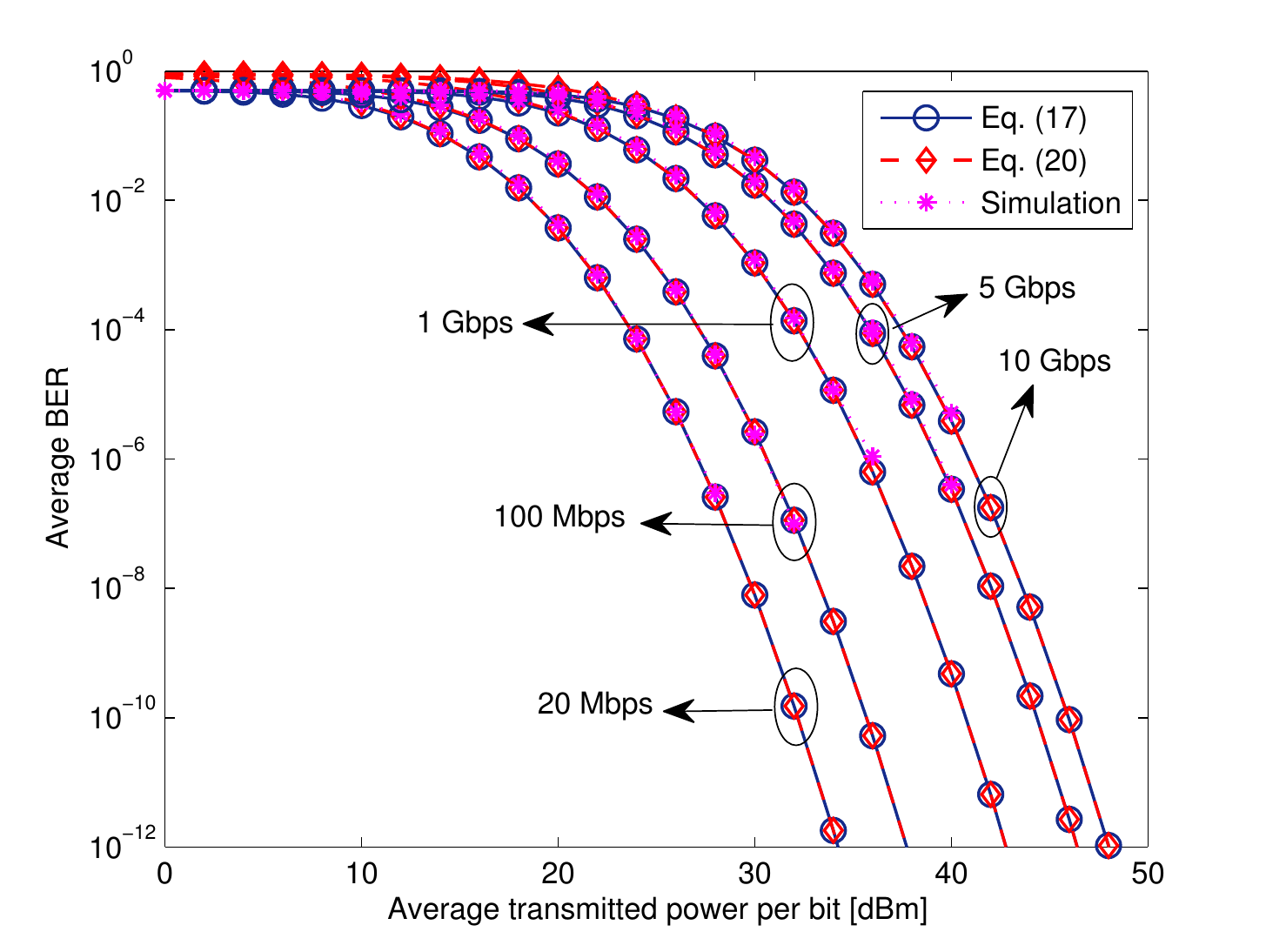}
       \caption {BER of a relay-assisted UWOC system with $N=3$ intermediate relay nodes and $105$ m end-to-end communication distance, for various data rates, i.e., $20$ Mbps, $100$ Mbps, $1$ Gbps, $5$ Gbps and $10$ Gbps.}
                   \vspace{-0.1in}
  \end{figure}
  \vspace{-0.04in}
  \section{Conclusion}
  In this paper, we analytically studied the end-to-end performance of UWOC systems with serial relaying. Our channel model is based upon the major degrading effects of the channel, namely absorption, scattering and turbulence-induced fading. Relying on Gauss-Hermite quadrature formula a closed-form solution for the BER of system under weak oceanic turbulence obtained. We also applied photon-counting method to evaluate the system BER in the presence of shot noise. Well matches between the results of analytical and photon-counting methods further confirmed the correctness of our assumption in derivation of analytical expressions, i.e., the negligibility of signal-dependent shot noise effect on the system performance. Moreover, excellent matches between analytical results and numerical simulations confirmed the accuracy of our derived expressions for the BER of multi-hop UWOC system with bit detect-and-forward relaying. Additionally, our results demonstrated that to reach a wide range underwater optical communication, designing the relay-assisted topology should be of utmost importance.
   For instance, dual-hop transmission in $45$ m  and $22.5$ m coastal waters, improved the system performance at the BER of $10^{-6}$ by $39$ dB and $17.5$ dB, respectively.

%
%


\end{document}